\documentclass{article}

\usepackage{spconf,amsmath,graphicx}
\usepackage{xcolor,soul}
\usepackage{booktabs} 
\usepackage{threeparttable}
\usepackage{cite}
\usepackage[hidelinks]{hyperref}
\usepackage{microtype}

\DeclareMathOperator*{\argmax}{arg\,max}

\title{End-to-end Lyrics Alignment for Polyphonic Music Using An Audio-to-Character Recognition Model}
\twoauthors
  {Daniel Stoller\sthanks{Work was conducted at Spotify.}}
	{Queen Mary University of London\\
	London, UK}
  {Simon Durand, Sebastian Ewert}
	{Spotify\\
	London, UK}

\begin{document}
\ninept

\renewcommand{\baselinestretch}{0.945}
\selectfont

\maketitle
\begin{abstract}
Time-aligned lyrics can enrich the music listening experience by enabling karaoke, text-based song retrieval and intra-song navigation, and other applications. Compared to text-to-speech alignment, lyrics alignment remains highly challenging, despite many attempts to combine numerous sub-modules including vocal separation and detection in an effort to break down the problem. Furthermore, training required fine-grained annotations to be available in some form. Here, we present a novel system based on a modified Wave-U-Net architecture, which predicts character probabilities directly from raw audio using learnt multi-scale representations of the various signal components. There are no sub-modules whose interdependencies need to be optimized. Our training procedure is designed to work with weak, line-level annotations available in the real world. With a mean alignment error of 0.35s on a standard dataset our system outperforms the state-of-the-art by an order of magnitude.
\end{abstract}
\section{Introduction}
\label{sec:intro}

Lyrics enable interacting with music in a plenitude of ways.
For example, one can search for songs if the title is unknown~\cite{Kruspe2016}, and time-aligned lyrics provide an intuitive way to navigate within a song, to sing along in a karaoke bar or to beep-out explicit content.
An early idea to enable such applications was to employ existing speech recognition methods to extract the lyrics from a given song~\cite{Mesaros2008}.
In practise, however, this had very limited success, as music might present the most challenging scenario such technologies could face.
Firstly, music often features many highly correlated sound sources, which strongly violates the assumption of statistical independence between the target and noise sources many speech recognition systems exploit~\cite{Barker2018}.
Secondly, while it is important for speech to be intelligible, singing voice is often used in creative ways and is more variable in the fundamental frequency range, timbre, tempo and the dynamics.
In fact, the task is so challenging that even humans frequently make mistakes\footnote{``misheard lyrics": \url{http://www.amiright.com/misheard/}}.

Fortunately, lyrics are often provided in textual form by various sources.
Therefore, the given lyrics-text only needs to be temporally aligned to a corresponding song to enable all of the above applications, which is a greatly simplified task.
Yet, even this simplified setting is highly challenging, with state of the art systems still yielding rather low alignment accuracy~\cite{MIREX2017Alignment}.
To improve the performance, many methods simplify the problem even further, e.g. by relying on lyrics being pre-aligned at a phrase level~\cite{Mesaros2010}. Other systems are designed for singing-only recordings, which often leads to a significant drop in performance when applied to polyphonic music~\cite{Kruspe2018}.
Additionally, many previous approaches rely on the availability of fine-grained ground-truth annotations during training -- since those are typically missing in practise, often complex and error-prone procedures iterating between re-training and re-alignment are employed~\cite{Gupta2018, Kruspe2016a}.

In this paper, we present a method employing a multi-scale neural network based on the Wave-U-Net architecture \cite{Stoller2018a} that predicts character probabilities end-to-end directly from raw audio -- in contrast to many previous approaches which often incorporate a wide range of different sub-modules, whose inter-dependencies are not easily optimized. Our system can instead learn to perform and combine such sub-tasks, including spectral front-ends and vocal processing techniques, as needed.
Our system can easily be trained using weak, line-level alignment annotations more readily available in the real-world -- in particular, it does not use any additional fine-grained annotations, which greatly simplifies the training process. 
As shown by our experiments, our system considerably improves the alignment accuracy over the state of the art on real-world polyphonic music (MIREX).
Further, combined with a simple language model, early experiments indicate that our acoustic model might even yield transcription results useful for various retrieval tasks in the future~\cite{Kruspe2018}.
Finally, we also provide a freely available dataset of 20 songs with a variety of genres from Jamendo\footnote{Available at \url{https://github.com/f90/jamendolyrics}} for the evaluation of alignment and transcription systems, to complement current evaluation datasets that are private and biased towards Pop songs~\cite{Mauch2010, MIREX2017Alignment}.

\section{Related work}

Given the close connection, various approaches for lyrics alignment employed speech processing methods in some form \cite{Fujihara2012}. In~\cite{Mesaros2010,Kruspe2016a,Kawai2017,Gupta2018}, pre-trained speech recognition models (phoneme detectors) are adapted in various ways as a work-around for the lack of accurately annotated music recordings.
However, only low accuracies were reported, which might be due to some properties differing considerably between singing and speech, including the syllable duration and word pronunciation.
Furthermore, the accompaniment typically adds a complex and structured source of noise to the problem, and often dominates the recording in terms of overall energy.
To circumvent this issue, many approaches~\cite{Gupta2018,Tsai2018,Kawai2017,Lee2017} operate only on solo singing recordings, which typically leads to accurate results -- the majority of (commercial) music, however, is polyphonic and thus the performance in this most common case either remains unknown for these methods~\cite{Gupta2018} or was shown to be significantly lower~\cite{Mesaros2008,Kruspe2016a,MIREX2017Alignment}.
For instance in~\cite{Kruspe2016a}, the mean absolute error on a-capella music is 0.67s, which rises to 10.14s on strongly polyphonic pieces~\cite{MIREX2017Alignment}.
To suppress the accompaniment, some approaches have thus employed singing voice separation techniques as a pre-processing step~\cite{Mesaros2008,Fujihara2011,Dzhambazov2015}. However, besides complicating the pipeline, this step so far tended to add separation artifacts that can render a phoneme unrecognizable~\cite{Dzhambazov2015}, and current state of the art systems require large additional training sets.

Several approaches make additional assumptions to further simplify the problem.
For example, the method presented in~\cite{Mauch2010} assumes that chord labels are attached to the lyrics and exploits them during the alignment process.
Other approaches assume that the lyrics are pre-aligned at a line or phrase level so that the method only needs to refine the alignment within these sections~\cite{Mesaros2010,Chien2016,Lee2017}.
Since music often contains repeated segments, some methods additionally analyze and compare the musical structure in a recording and in  corresponding lyrics~\cite{WangKNSY04_LyricAlly_ACMMM,McVicarEG2014_RepetitionLyricsAlign_ICASSP}.

Furthermore, many systems rely on rather complex training or parameter optimization procedures, which can affect the training duration or reliability. For example, the phoneme detectors mentioned above require a fine-grained phoneme labelling during training. As such a dataset is not available for music, the system in~\cite{Kruspe2016a} periodically re-calculates an alignment between the lyrics and recordings in the training dataset (Viterbi forced alignment) and continues a frame-wise training based on the results. This procedure is a variant of Viterbi training~\cite{FranziniLW1990_ConnectionistViterbiTraining_ICASSP}, which was found to accelerate convergence in some cases but which often also led to inferior model performance as the hard-alignment can bias the training towards solutions that generalize less well compared to approaches using soft-alignments~(Baum-Welch training)~\cite{DurbinEKM98_BioSequenceAnalysis_BOOK}.
Finally, systems often consist of multiple complex stages~\cite{Fujihara2012,Fujihara2011,Chien2016,Lee2017,Tsai2018}, introducing many parameters that are not optimized jointly, so that errors tend to propagate between stages.
In contrast, all parameters in our system are trained jointly on polyphonic music, we only require weak alignment annotations on the level of lyrical lines and employ a ``soft-alignment" during training to stabilize the model performance.

\section{Proposed Method}

The central component in our alignment method is an acoustic model which, given a partition of a music recording into short slices of time, defines for each slice a probability distribution over all characters we could observe. More precisely, given a monaural audio signal $x \in [-1,1]^I$ with $I$ input samples, our goal is to estimate probabilities $P \in [0,1]^{T \times |\hat{\mathcal{C}}|}$ of the characters used in the lyrics. Here, we define $\mathcal{C} := \{ a, b, \ldots, z, ', \sqcup  \}$ as the set of characters our model supports, and $\hat{\mathcal{C}} := \mathcal{C}\ \cup \{\epsilon\}$ additionally contains the blank symbol $\epsilon$. This symbol will enable us to perform non-linear time-warpings during the subsequent alignment step, as we will see later. Note that each of the $T$ time slices corresponds to $\lfloor I/T\rfloor$ samples.

\subsection{Acoustic model}
\label{sec:AcousticModel}

\begin{figure}[tb]

\begin{minipage}[b]{1.0\linewidth}
  \centering
  \centerline{\includegraphics[width=8.5cm]{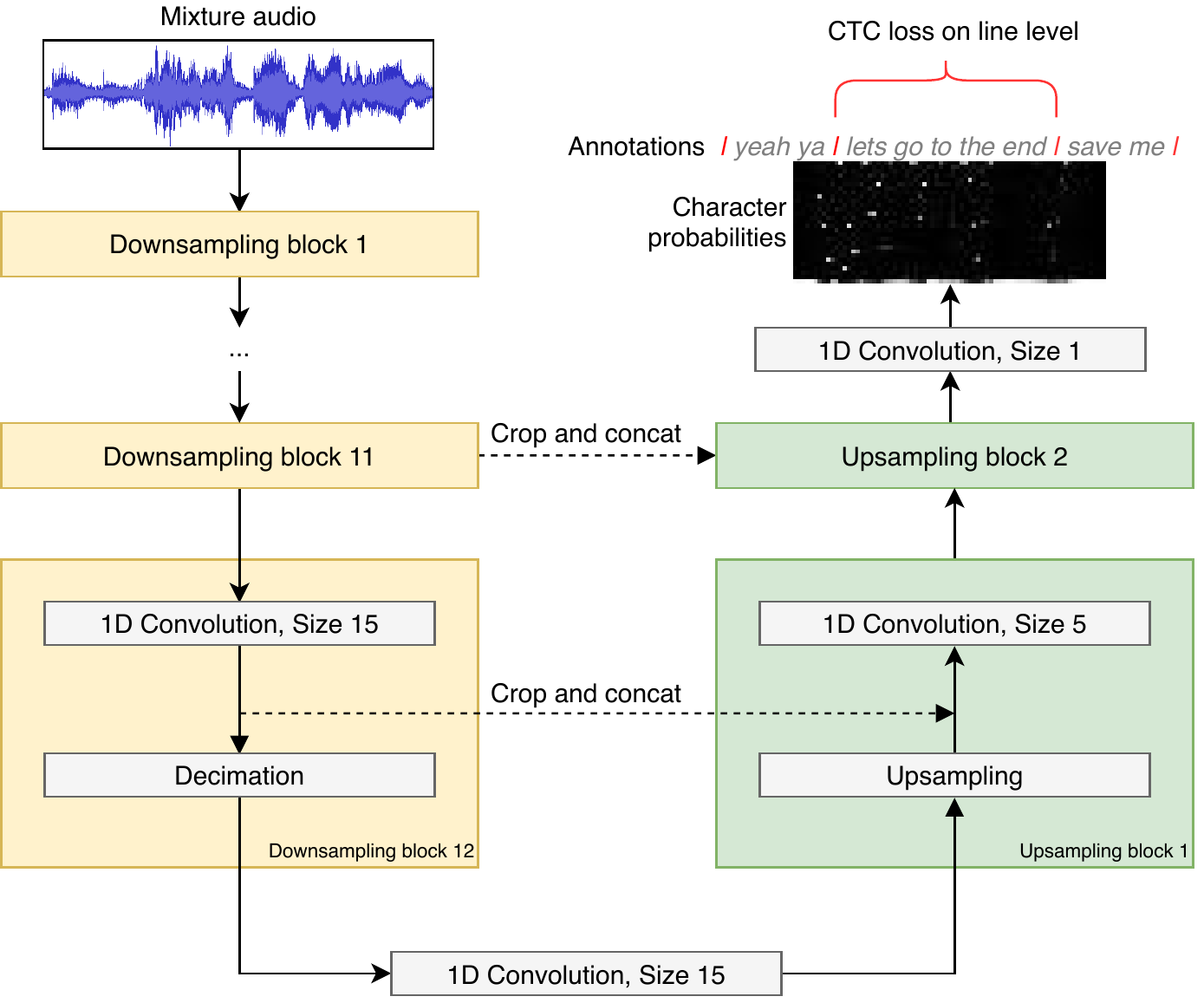}}
\end{minipage}
\caption{Our proposed lyrics transcription and alignment model adapted from the Wave-U-Net. The CTC loss is applied on lyrical lines fully contained within the output window}
\label{fig:model}
\end{figure}

For the acoustic model $f_{\theta}$, we adapt the architecture of the Wave-U-Net (variant M4)~\cite{Stoller2018a}.
The model was originally proposed for singing voice separation and was designed to model highly non-stationary vocals as well as the accompaniment. Inspired by wavelets, it constructs signal representations at multiple time resolutions -- which suggests that it can not only model sound sources at different time-frequency resolutions, but also capture low-level phoneme as well as high-level word-based information to inform character-based lyrics prediction.

We adapt the model as shown in Figure~\ref{fig:model}. Similar to the original Wave-U-Net, our network employs a series of blocks combining a 1D-convolution and a downsampling layer, whose receptive field grows exponentially with the number of layers and thus enables an efficient translation of low-level into higher level features. However, compared to a regular convolutional network architecture~\cite{KrizhevskySH2012_Alexnet_NIPS}, we do not stop after reaching the time-resolution required for the output, but continue with further downsampling steps to obtain even higher level features. To yield the desired output, we add upsampling layers, which increase the time resolution again.
The model concatenates the output of previous downsampling layers with the input to upsampling layers -- this way, the lower layers can focus on representing higher level features and do not need to encode fine-grained information, which is crucial for performance.

Overall, we set the input size to $352243$ audio samples ($15.97$s at $22.05$ KHz) and first process them with 12 downsampling blocks. We then use two upsampling blocks resulting in about 20 character probability distributions per second, which enables a temporally precise alignment.
We also use a context-aware prediction framework (see \cite{Stoller2018a} for additional details) such that the output predictions do not cover the entire input but the centre $225501$ samples ($10.23$s). For song-wise predictions, the model is thus shifted in multiples of $225501$ samples across the song, and character predictions are concatenated to a single probability matrix $P$.

\subsection{Marginalising over possible alignments during training}

Using frame-level character labels, it would be straightforward to train $f_{\theta}$ in a supervised fashion as a classifier, i.e.\,typically using a cross-entropy loss. Unfortunately, a dataset containing such detailed labels is not available for polyphonic music. To enable the use of weakly aligned lyrics data, the methods in~\cite{Kruspe2016a,Gupta2018} employ a procedure resembling Viterbi training~\cite{FranziniLW1990_ConnectionistViterbiTraining_ICASSP}, where intermediate models are employed to force-align the lyrics to corresponding audio recordings. Frame-level annotations are then generated from the aligned lyrics. Viterbi training was found to accelerate convergence in some cases \cite{FranziniLW1990_ConnectionistViterbiTraining_ICASSP} but can also lead to inferior model performance as the hard-alignment tends to bias the training towards solutions that generalize less well, see e.g.\,~\cite[Chapter~6]{DurbinEKM98_BioSequenceAnalysis_BOOK}.

The connectionist temporal classification (CTC)~\cite{Graves2006} loss offers an alternative to Viterbi training and is used today in many state of the art speech recognition systems in some form~\cite{Barker2018}. The CTC loss is essentially a simplified version of the forward-backward procedure used to calculate the posterior marginals of the states in a hidden Markov model (HMM)~\cite{Rabiner1989_HMMTutorial_IEEE}, and can only be applied to left-to-right~(Bakis-type) HMMs with uniform transition probabilities~\cite{Rabiner1989_HMMTutorial_IEEE}. 
The CTC loss takes the character probability distributions $P$ generated by the acoustic model $f_{\theta}$ to calculate a ``soft-alignment" between the time slices in $P$ and each character in a given target sequence $y$, which can be represented as a probability distribution over all possible alignments.
Similar to Gaussian mixture model training, we can marginalise over this distribution to obtain the likelihood of the target sequence  
\begin{equation}
    p(y|x) = \sum_{\hat{y} \in \hat{\mathcal{C}}^T, B(\hat{y}) = y}\ \prod_{t=1}^T P_{t, \hat{y}_t}
    \label{eq:ctc_loss}
\end{equation}
and use it for maximum likelihood-based training of the acoustic model.
The operator $B(\cdot)$ takes a sequence $\hat{y} \in \hat{\mathcal{C}}^T$ and removes repeated symbols and blanks, see \cite{Graves2006} for further algorithmic details. 
In contrast to Viterbi training that picks the most likely alignment, we marginalize over all possible alignments between $x$ and $y$, using a sum operation instead of a max operation in~\eqref{eq:ctc_loss}.

To use this loss directly, however, the model would need to make predictions for the entire song $x$.
This creates memory issues in practice due to the long length of the sequences $x$, and does not exploit the line-level alignments available.
We thus calculate character probabilities for a chunk of audio with a fixed size, and apply the CTC loss with individual lyrical lines as target sequences, using only slices of $P$ corresponding to a time position between the start and end times of the lyrical line.

\begin{figure}[t]
\begin{minipage}[b]{1.0\linewidth}
  \centering
  \centerline{\includegraphics[width=8.5cm]{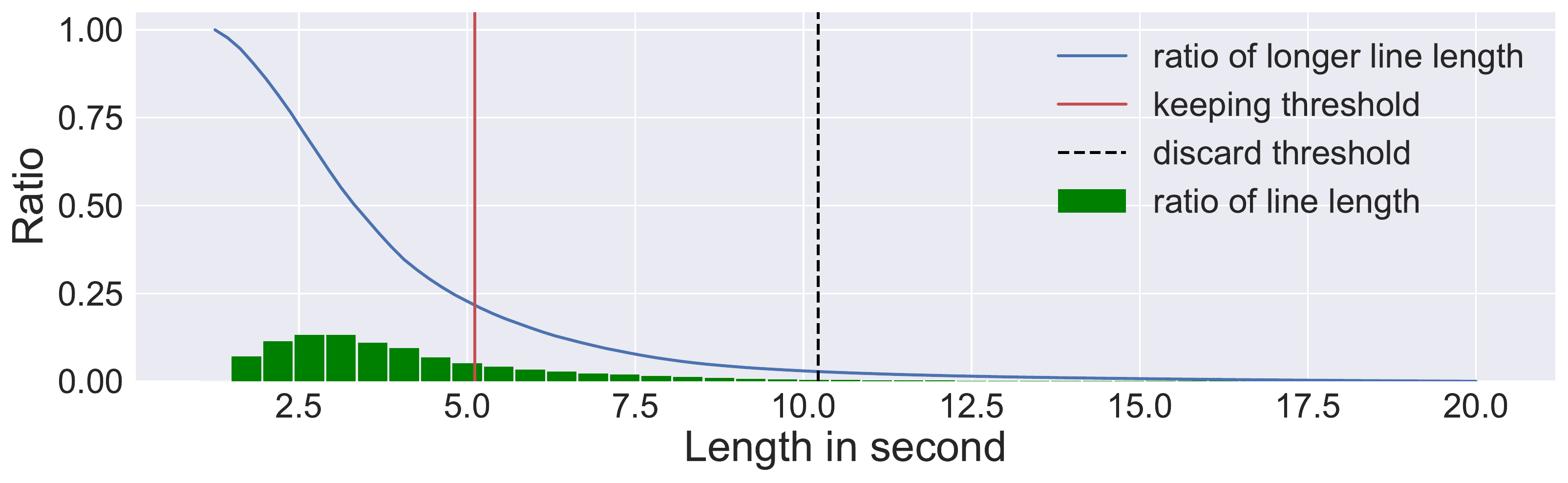}}
\end{minipage}
\caption{Histogram of line lengths in the training set (green bars). The blue line shows the relative number of lines that are longer than a given duration. All the lines shorter than the value indicated by the red vertical line are kept, and longer than the value indicated by the black dotted vertical line discarded.}
\label{fig:distrib}
\end{figure}

Using the CTC loss influenced our design choices for the acoustic model in several ways. 
First, one reason for selecting an output size of $10.23$ seconds (compare Section~\ref{sec:AcousticModel}) is that most lyrical lines fit into such a window, compare also Figure~\ref{fig:distrib} which shows the distribution of the length of lines in the training set.
Second, we use a purely convolutional network (with a large input context) and do not add recurrent connections.
This is inspired by the findings in \cite{Sak2015}, where the authors demonstrate that while RNNs in combination with a CTC loss often produce good speech recognition results, they can suffer from low temporal accuracy when used for alignment. The authors argue that the CTC loss does not favour any particular kind of alignment and thus an RNN might learn to wait for more inputs before emitting symbols to reduce the uncertainty.
Based on these results, we decided to explore a convolutional architecture and indeed we did not observe a similar behaviour in our experiments.

\subsection{Alignment Procedure}

Given a music recording $x$ and corresponding lyrics $y$, we employ the trained acoustic model in a procedure resembling Viterbi forced-alignment.
As a first step, we pre-process the lyrics by removing all symbols not supported by the acoustic model.
Assuming $x$ contains $T$ time slices, our goal is to find an aligned sequence $\tilde{y} \in \hat{\mathcal{C}}^T$ with maximum probability under the acoustic model predictions $P$ whose corresponding reduced form~$B(\tilde{y})$ is equal to the given lyrics~$y$. More precisely:
\begin{equation}
\tilde{y} := \argmax_{\hat{y} \in \hat{\mathcal{C}}^T, B(\hat{y}) = y}\ \prod_{t=1}^T P_{t, \hat{y}_t}
\end{equation}
In other words, the alignment is typically encoded by inserting blank symbols $\epsilon$ into $y$ such that the resulting longer sequence $\tilde{y}$ runs synchronously to the audio.
Relying on dynamic programming, we can compute $\tilde{y}$ with a time complexity of $\mathcal{O}(TO)$, where $O$ is the length of~$y$~\cite{Rabiner1989_HMMTutorial_IEEE}.
As commonly done, the probability products are calculated in the log-domain to avoid numerical instabilities.
To prevent the rare case that $P_{t, c} = 0\ \forall t \in \{1, \ldots, T\}$ for a symbol $c \in \mathcal{C}$ appearing in the lyrics, resulting in all alignments $\hat{y}$ having an assigned probability of zero, we add a small amount of uniformly distributed noise sampled from $U[10^{-11}, 10^{-10}]$ to all entries in $P$.
Finally, we add a constant delay of 180ms to the alignment, which we chose to optimize performance on a small validation dataset.

\subsection{Using the Acoustic Model for Transcription}
\label{sec:language_model}

For transcription, we aim to find the most likely output sequence $\argmax_y p(y|x)$ for a given input $x$.
Since this is not computationally feasible, we approximate the solution by beam search decoding, using a beam width of $1024$.
Since it is difficult for the acoustic model to learn word spelling in addition to acoustic recognition, we also perform decoding weighted with a tri-gram, word-level language model trained on lyrics text\footnote{\url{https://github.com/parlance/ctcdecode}}.
The language model weight $\alpha$ and word insertion penalty $\beta$ are optimised for WER on the \textit{Mauch} dataset~\cite{Mauch2010} with a grid search over $\{0.0, 0.2,\ldots,2.0\}$, obtaining $\alpha=0.2$ and $\beta=0.4$ for evaluation on the Jamendo dataset.

\section{Experiments}

\subsection{Experimental setup}

\subsubsection{Dataset}

For training, we use an internal dataset comprised of 44,232 songs with English lyrics and varied Western genres such as Pop, Rock and Hip-Hop.
The songs are annotated with start and end times of each lyrical line, and the distribution of their lengths is shown in Figure~\ref{fig:distrib}.
We use 39232 songs for training and 5000 for validation.

We convert all songs to mono signals sampled at $22.05$ KHz.
For each song, we generate training samples of the required length, $15.97$s, by shifting across the input in increments of half of the duration covered by the model's character predictions, determining which lyrical lines start and end within the model's output window, and generating a training sample for each.
This is so that all lyrical lines shorter than $\frac{10.22}{2} = 5.11$ seconds and most up to $10.22$ seconds are included, which make up $78.4$\% and $97.2$\% percent of all lyrical lines, respectively, as seen in Figure~\ref{fig:distrib}.
If there are no lines overlapping with the output window, an empty label is generated to ensure the model predicts silence by outputting the blank symbol $\epsilon$ for instrumental sections.
The above procedure ensures a mostly uniform sampling from the audio signals and avoids potential model biases with audio examples synchronised to lyrical lines.
Since we restrict our model's output vocabulary to the English alphabet (only lower-case), a whitespace and an apostrophe, we convert our lyrics labels to lower-case and remove other unsupported symbols.

\subsubsection{Training procedure}

We optimise the CTC loss using the ADAM Optimiser (learning rate~$10^{-4}$, $\beta_1 = 0.9$, $\beta_2 = 0.999$, $\epsilon = 10^{-8}$) and a batch size of $32$.
Every 10,000 iterations, we compute the training loss averaged over the last 10,000 iterations.
After 6 successive times without improvement, we reduce the learning rate to $10^{-5}$, and continue training, again until we do not see improvement 6 times.
Finally, we select the model with the best performance on the validation set.
The training time was 25 hours using 4 NVIDIA Tesla P100 GPUs.

\subsection{Alignment Results}
\label{sec:alignment_results}
    
    \begin{table}[t!]
    \footnotesize
    \centering
    \setlength{\tabcolsep}{1.2ex}
    \begin{tabular}{cccccccc}
    \toprule
    & \multicolumn{6}{c}{Mauch} & \multicolumn{1}{c}{Jamendo} \\
    \cmidrule(lr){2-7} \cmidrule(lr){8-8}
    Metric & AK1 & AK2 & AK3 & DMS1 & DMS2 & Ours & Ours \\
    \midrule
    AE & 17.70 & 22.23 & 9.03 & 14.91 & 11.64 & \textbf{0.35} & \textbf{0.82} \\
    Perc & 8.5 & 2.4 & 15.4 & 3.8 & 13.8 & \textbf{77.2} & \textbf{70.4} \\
    \bottomrule
    \end{tabular}
    \caption{Alignment accuracy compared to systems evaluated for MIREX 2017~\cite{MIREX2017Alignment} and on our Jamendo dataset (see Section~\ref{sec:alignment_results}).}
    \label{tab:mauch_results}
    \vspace{-0.2cm}
    \end{table}

We evaluated our alignment system on the \textit{Mauch} dataset~\cite{Mauch2010}. Since it was also used in the MIREX 2017 lyrics alignment challenge, we can compare our results with all approaches submitted to the challenge (see Table~\ref{tab:mauch_results}).
\textit{AE} is the mean absolute deviation in seconds from the predicted to the true word start times, averaged over songs, and 
\textit{Perc} the average percentage of time in a song that the predicted position in the lyrics is correct (see Fig. 9,~\cite{Fujihara2011}).
We see that our method vastly outperforms all others, predicting the currently sung word $77.2$\% of the time, an absolute $61.8$\% higher than the best MIREX method (AK3).

However, since the Mauch dataset contains only Pop music, the performance in many real-world scenarios might differ.
We thus built the \textit{Jamendo} dataset, containing 20 songs from nine genres, with annotations of start and end times for all words in the lyrics.
In contrast to previously used datasets~\cite{Mauch2010,MIREX2017Alignment}, we make it freely available to enable a straightforward, comparable evaluation across approaches.
Compared to the Mauch dataset, the AE on Jamendo increases moderately with our approach, as shown in Table~\ref{tab:mauch_results}.
We find that the median of absolute errors does not increase however, indicating more extreme prediction outliers, caused by some Hip-Hop and Metal songs, likely due to slurred pronunciation.
However, performance overall remains strong despite the increase in genre diversity.

\subsection{Transcription Results}
\label{sec:transcription_results}

We evaluated the transcription performance using character error rate (CER) and word error rate (WER) on the Mauch and Jamendo datasets, and compare between beam search and LM decoding. 
The results shown in Table~\ref{tab:transcription_results} show a CER of around 50\% across the board.
However, we find that words are often slightly misspelt, likely due to the model needing to learn spelling in addition to acoustic recognition without a vocabulary, and as a result the WER is considerably higher.
However, the use of the language model decoding from Section~\ref{sec:language_model} helps and improves WER significantly, while keeping CER mostly constant.

    \begin{table}[t]
        \footnotesize
        \centering
        \begin{threeparttable}
        \begin{tabular}{cccccc}
        \toprule
        & & \multicolumn{2}{c}{Mauch} & \multicolumn{2}{c}{Jamendo} \\
        \cmidrule(lr){3-4} \cmidrule(lr){5-6}
        Model & Decoder & WER & CER & WER & CER \\
        \midrule
        Ours & Beam & 80.4 & \textbf{48.9} & 84.4 & \textbf{49.2} \\
        Ours & LM & \textbf{70.9}* & 49.4* & \textbf{77.8} & 50.2 \\
        \bottomrule
        \end{tabular}
        
        \begin{tablenotes}
            \footnotesize
            \item[*]{after optimising the language model on Mauch dataset}
        \end{tablenotes}
        \end{threeparttable}
    \caption{Transcription accuracy of our approach on the Mauch and Jamendo datasets (see Section~\ref{sec:transcription_results}).}
    \label{tab:transcription_results}
        
    \end{table}

\section{Discussion}

In contrast to previous work employing many separate stages such as pre-processing, voice separation and detection~\cite{Fujihara2012}, we jointly optimise all model parameters and achieve an alignment accuracy considerably above the state of the art.
As we directly output character probabilities and not phonemes as commonly done~\cite{Fujihara2012}, we do not have to convert the lyrics into phonemes using a pronunciation dictionary.
Since these are usually built for speech and assume that every word has exactly one pronunciation, they are less suitable for lyrics due to the way pronunciation is often extensively varied in singing voice and since they do not contain rules for vocalisations such as ``aah" and ``ooh".
However, phoneme-based models might be more easily adapted to other languages for the alignment task, since only the phoneme dictionary has to be replaced.
Further, while the model was designed for alignment and not transcription, the results are already useful for a range of retrieval applications.

Finally, we investigated the impact the presence of accompaniment has on our model's performance. To this end, we employed a large internal dataset to re-train the vocal separation method presented in~\cite{Jansson2017} and used it to extract the vocals from each dataset. We then re-trained our acoustic model on the extracted vocals. This way, we were able to further lower the AE from $0.82$ to $0.38$ (Perc up from $70.4$\% to $76.8$\%) on our Jamendo dataset, and from $0.35$ to $0.27$ (Perc up from $77.2$\% to $78.1$\%) on the Mauch dataset. Since the separator was not optimised directly for lyrics alignment or transcription however, these results suggest potential avenues for multi-task learning in future work to exploit multiple datasets and unify different training objectives -- with the drawback that such large datasets need to be available.

\section{Conclusion}

We proposed a modified Wave-U-Net architecture that employs learnt multi-scale representations to predict character probabilities directly from the waveform of polyphonic music. In contrast to most existing systems, the system can be trained end-to-end to avoid complex non-optimized component inter-dependencies and requires only weak, line-level alignment annotations during training.
Applied to lyrics alignment, our system considerably outperformed state of the art systems, which were evaluated for the MIREX lyrics alignment challenge. Used for lyrics transcription, the system achieves a performance enabling various retrieval tasks -- despite not being designed for transcription. 
Furthermore, we make an annotated dataset freely available to support the evaluation of future lyrics alignment and transcription systems and to encourage comparability of results and research in a more realistic and challenging setting.

\vspace{\baselineskip}
\noindent \textbf{Acknowledgments:} We thank Georgi Dzhambazov for assisting with the evaluation in the context of the MIREX challenge.

\bibliographystyle{IEEEbib}
\bibliography{main}

\begin{thebibliography}{10}

\bibitem{Kruspe2016}
Anna~M Kruspe,
\newblock ``Retrieval of textual song lyrics from sung inputs,''
\newblock in {\em Proceedings INTERSPEECH}, 2016, pp. 2140--2144.

\bibitem{Mesaros2008}
Annamaria Mesaros and Tuomas Virtanen,
\newblock ``Automatic alignment of music audio and lyrics,''
\newblock in {\em Proceedings of the International Conference on Digital Audio
  Effects (DAFx)}, 2008.

\bibitem{Barker2018}
Jon Barker, Shinji Watanabe, Emmanuel Vincent, and Jan Trmal,
\newblock ``The fifth {CHiME} speech separation and recognition challenge:
  Dataset, task and baselines,''
\newblock in {\em Proceedings Interspeech}, 2018.

\bibitem{MIREX2017Alignment}
MIREX,
\newblock ``Lyrics alignment results,''
  \url{https://www.music-ir.org/mirex/wiki/2017:Automatic_Lyrics-to-Audio_Alignment_Results},
  2017.

\bibitem{Mesaros2010}
Annamaria Mesaros and Tuomas Virtanen,
\newblock ``Automatic recognition of lyrics in singing,''
\newblock {\em EURASIP Journal on Audio, Speech, and Music Processing}, vol.
  2010, no. 1, pp. 1, 2010.

\bibitem{Kruspe2018}
Anna~M Kruspe,
\newblock {\em Application of automatic speech recognition technologies to
  singing},
\newblock Ph.D. thesis, {{Technische Universit{\"a}t Ilmenau}}, 2018.

\bibitem{Gupta2018}
Chitralekha Gupta, Rong Tong, Haizhou Li, and Ye~Wang,
\newblock ``Semi-supervised lyrics and solo-singing alignment,''
\newblock in {\em Proceedings of the International Society for Music
  Information Retrieval Conference ({ISMIR})}, 2018, pp. 600--607.

\bibitem{Kruspe2016a}
Anna~M Kruspe,
\newblock ``Bootstrapping a system for phoneme recognition and keyword spotting
  in unaccompanied singing,''
\newblock in {\em Proceedings of the International Conference on Music
  Information Retrieval (ISMIR)}, 2016, pp. 358--364.

\bibitem{Stoller2018a}
Daniel Stoller, Sebastian Ewert, and Simon Dixon,
\newblock ``{Wave-U-Net}: A multi-scale neural network for end-to-end source
  separation,''
\newblock in {\em Proceedings of the International Society for Music
  Information Retrieval Conference ({ISMIR})}, 2018, vol.~19, pp. 334--340.

\bibitem{Mauch2010}
Matthias Mauch, Hiromasa Fujihara, and Masataka Goto,
\newblock ``Lyrics-to-audio alignment and phrase-level segmentation using
  incomplete internet-style chord annotations,''
\newblock in {\em Proceedings of the Sound Music Computing Conference (SMC)},
  2010, pp. 9--16.

\bibitem{Fujihara2012}
Hiromasa Fujihara and Masataka Goto,
\newblock ``Lyrics-to-audio alignment and its application,''
\newblock in {\em Dagstuhl Follow-Ups}. Schloss Dagstuhl-Leibniz-Zentrum fuer
  Informatik, 2012, vol.~3.

\bibitem{Kawai2017}
Dairoku Kawai, Kazumasa Yamamoto, and Seiichi Nakagawa,
\newblock ``Lyric recognition in monophonic singing using pitch-dependent
  {DNN},''
\newblock in {\em Proceedings of the {IEEE} International Conference on
  Acoustics, Speech and Signal Processing ({ICASSP})}, 2017, pp. 326--330.

\bibitem{Tsai2018}
Che-Ping Tsai, Yi-Lin Tuan, and Lin-shan Lee,
\newblock ``Transcribing lyrics from commercial song audio: the first step
  towards singing content processing,''
\newblock in {\em Proceedings of the IEEE International Conference on
  Acoustics, Speech and Signal Processing (ICASSP)}, 2018, pp. 5749--5753.

\bibitem{Lee2017}
Sang~Won Lee and Jeffrey Scott,
\newblock ``Word level lyrics-audio synchronization using separated vocals,''
\newblock in {\em Proceedings of the {IEEE} International Conference on
  Acoustics, Speech and Signal Processing ({ICASSP})}, 2017, pp. 646--650.

\bibitem{Fujihara2011}
Hiromasa Fujihara, Masataka Goto, Jun Ogata, and Hiroshi~G Okuno,
\newblock ``Lyricsynchronizer: Automatic synchronization system between musical
  audio signals and lyrics,''
\newblock {\em IEEE Journal of Selected Topics in Signal Processing}, vol. 5,
  no. 6, pp. 1252--1261, 2011.

\bibitem{Dzhambazov2015}
Georgi Dzhambazov and Xavier Serra,
\newblock ``Modeling of phoneme durations for alignment between polyphonic
  audio and lyrics,''
\newblock in {\em Proceedings of the Sound and Music Computing Conference
  (SMC)}, 2015.

\bibitem{Chien2016}
Yu-Ren Chien, Hsin-Min Wang, and Shyh-Kang Jeng,
\newblock ``Alignment of lyrics with accompanied singing audio based on
  acoustic-phonetic vowel likelihood modeling,''
\newblock {\em IEEE/ACM Transactions on Audio, Speech, and Language
  Processing}, vol. 24, no. 11, pp. 1998--2008, 2016.

\bibitem{WangKNSY04_LyricAlly_ACMMM}
Ye~Wang, Min-Yen Kan, Tin~Lay Nwe, Arun Shenoy, and Jun Yin,
\newblock ``{LyricAlly}: automatic synchronization of acoustic musical signals
  and textual lyrics,''
\newblock in {\em Proceedings of the Annual ACM International Conference on
  Multimedia (ACMMM)}, 2004, pp. 212--219.

\bibitem{McVicarEG2014_RepetitionLyricsAlign_ICASSP}
Matt McVicar, Daniel~PW Ellis, and Masataka Goto,
\newblock ``Leveraging repetition for improved automatic lyric transcription in
  popular music,''
\newblock in {\em Proceedings of the IEEE International Conference on
  Acoustics, Speech and Signal Processing (ICASSP)}, 2014, pp. 3117--3121.

\bibitem{FranziniLW1990_ConnectionistViterbiTraining_ICASSP}
Michael Franzini, K-F Lee, and Alex Waibel,
\newblock ``Connectionist {V}iterbi training: a new hybrid method for
  continuous speech recognition,''
\newblock in {\em Proceedings of the IEEE International Conference on
  Acoustics, Speech and Signal Processing (ICASSP)}, 1990, pp. 425--428.

\bibitem{DurbinEKM98_BioSequenceAnalysis_BOOK}
Richard Durbin, Sean~R Eddy, Anders Krogh, and Graeme Mitchison,
\newblock {\em Biological sequence analysis: probabilistic models of proteins
  and nucleic acids},
\newblock Cambridge university press, 1998.

\bibitem{KrizhevskySH2012_Alexnet_NIPS}
Alex Krizhevsky, Ilya Sutskever, and Geoffrey~E Hinton,
\newblock ``Imagenet classification with deep convolutional neural networks,''
\newblock in {\em Advances in Neural Information Processing Systems}, 2012, pp.
  1097--1105.

\bibitem{Graves2006}
Alex Graves, Santiago Fern{\'a}ndez, Faustino Gomez, and J{\"u}rgen
  Schmidhuber,
\newblock ``Connectionist temporal classification: labelling unsegmented
  sequence data with recurrent neural networks,''
\newblock in {\em Proceedings of the ACM International Conference on Machine
  learning (ICML)}, 2006, pp. 369--376.

\bibitem{Rabiner1989_HMMTutorial_IEEE}
Lawrence~R Rabiner,
\newblock ``A tutorial on hidden {M}arkov models and selected applications in
  speech recognition,''
\newblock {\em Proceedings of the IEEE}, vol. 77, no. 2, pp. 257--286, 1989.

\bibitem{Sak2015}
Ha{\c{s}}im Sak, Andrew Senior, Kanishka Rao, Ozan Irsoy, Alex Graves,
  Fran{\c{c}}oise Beaufays, and Johan Schalkwyk,
\newblock ``Learning acoustic frame labeling for speech recognition with
  recurrent neural networks,''
\newblock in {\em Proceedings of the IEEE International Conference on
  Acoustics, Speech and Signal Processing (ICASSP)}, 2015, pp. 4280--4284.

\bibitem{Jansson2017}
Andreas Jansson, Eric~J. Humphrey, Nicola Montecchio, Rachel Bittner, Aparna
  Kumar, and Tillman Weyde,
\newblock ``Singing voice separation with deep {U-Net} convolutional
  networks,''
\newblock in {\em Proceedings of the International Society for Music
  Information Retrieval Conference (ISMIR)}, 2017, pp. 323--332.

\end{thebibliography}

\end{document}